\documentclass[10pt]{iopart}
\usepackage{graphicx}  
\usepackage{times}

\begin{document}
\renewcommand{\textheight}{7.48truein}

\newcommand{\name}[1]{#1}
\newcommand{\subsubsec}[1]{\subsubsection{#1}\quad\par\smallskip}
\newcommand{\QUERY}[1] {{}}
\newcommand{\OMIT}[1] {{}}
\newcommand{\quasicrystal}{quasicrystal }
\newcommand{\quasicrystals}{quasicrystals }

\title
[Clusters, phasons, and stabilisation]
{Discussion on clusters, phasons, and quasicrystal stabilisation}

\author{C.~L.~HENLEY\dag, M.~de~BOISSIEU\ddag, and W.~STEURER\S}

\address{\dag\ Dept. of Physics, Cornell University,
Ithaca NY 14853-2501, USA}

\address{\ddag\ Labo. de Thermodynamique et Physico Chimie M\'etallurgique,
UMR CNRS 5614, INPG, ENSEEG, BP 75, 38402 St. Martin d'H\`eres Cedex.}

\address{\S Laboratory of Crystallography, Dept. of Materials,
ETH Z\"urich, Wolfgang-Pauli-Strasse 10, 8093 Z\"urich, Switzerland}

\begin{abstract}
This paper summarises a two-hour discussion 
at the Ninth International Conference on Quasicrystals,
including nearly 20 written comments sent afterwards,
concerning (i) the meaning [if any] of clusters in  quasicrystals;
(ii) phason elasticity, and (iii) thermodynamic stabilisation
of quasicrystals.  
\end{abstract}

This paper represents a two-hour discussion 
that we moderated
at the Ninth International Conference on Quasicrystals.
The discussion was divided into three parts, concerning
(i) the meaning [if any] of clusters in  quasicrystals;
(ii) phason elasticity, and (iii) thermodynamic stabilisation
of quasicrystals.  
To prime the participants, the discussion leaders
prepared discussion papers in advance, 
which are printed separately~\cite{steurer-qdisc,deboissieu-qdisc,henley-qdisc}.

The colloquy was rather lively, considering that the
topics are probably the most intractable questions in quasicrystals
and the ones which have seen the least progress since 1985.
There was somewhat less detailed discussion of specific experiments
[done, or proposed] than might have been wished, but the discussion
made absolutely clear that more experiments are needed in each
of the three areas, if we are ever to resolve these questions.

Included in this text are (i) notes from the oral comments taken
at the discussion session; and (ii) written comments  -- about
18, almost evenly divided between theorists and experimentalists --
which were sent afterwords by most of the participants who spoke
during the session:
\name{Shelomo I. Ben-Abraham, Luc Barbier, Tomonari Dotera, Keiichi ~Edagawa, 
Franz ~G\"ahler,
Yuri ~Grin, Yasushi Ishii, Kaoru Kimura, Reinhard L\"uck, 
Ronan McGrath, Marek~Mihalkovi\v{c},
N.~K.~Mukhopadhyay, Gerardo ~Naumis, David  Rabson, Johannes Roth, 
Hans-Reiner Trebin,
Michael Widom, and Eeuwe Zijlstra}, and 
additionally Zorka Papadopoulos, who was not present at the discussion.
Besides the above, John Cahn, Denis Gratias, Ken F. Kelton,
and Patricia A. Thiel also made remarks.
The order has been rearranged freely so as to
gather related points in the same place, 
and the contributions were intensively edited
with the aim of improving clarity.

\QUERY{Comments in slant text -- often in ALL CAPITALS and/or
introduced with Q -- are my queries to authors, which must
be resolved or cut.}

\section{Clusters in quasicrystals}

{\it This discussion was led by \name{Walter Steurer}.}

The reader may be interested in \name{Eiji Abe's}
mini-review paper~\cite{e-abe}, which
forcefully puts forward a viewpoint on clusters that was
not strongly represented in this discussion (which he was
unable to attend); in effect, it should be considered as if
it were a fourth discussion leader's paper
alongside references \cite{steurer-qdisc}, \cite{deboissieu-qdisc}, 
and \cite{henley-qdisc}.

\subsection{Defining `clusters'}

The first (and longest) question was the definition of `cluster'.
Some participants (including \name{Steurer})
thought it was valuable to arrive at an agreed definition.
[\name{Mihalkovi\v{c}} recalled that in the old days,
the community succeeded on one clarification of nomenclature:
the distinction between `packing' and `matching' rules.]
But others in the discussion took
a libertarian attitude that `cluster' can mean anything, so long
as the user prefaces his or her use by specifying what sense
is meant, e.g.  \name{Gratias} said, Let's each define our own
definition of a `cluster', with a meaning limited to that talk or paper.
\name{G\"ahler} replied,  it's just important that each time we
use it, we say  what we mean.
\name{Mihalkovi\v{c}} observed that 
the question of the physical reality of clusters is different 
from the question of synchronizing our dictionaries;
we should {\it not} focus on finding the 
`best' definition for `cluster',  but merely on 
eliminating any clearly misleading concepts or terms.

The  main tension was between geometrical definitions -- based on
purely structural data -- and chemical definitions including
some criterion about bonding.  The practicality of a bonding
criterion tends to hang, of course, on the experimental
question whether strongly-bonded atom clusters actually exist
in \quasicrystals, which is discussed later in three
micro-reviews (subsections~\ref{sec:cluster-Kimura},\ref{sec:cluster-McGrath}, 
and \ref{sec:cluster-surface-Barbier}.  

For some participants, {\it non-overlapping}
is an automatic attribute of a `cluster',
yet for others, certain `clusters' are of interest precisely
because they completely -- and overlappingly -- {\it cover} the
whole \quasicrystal structure.
Some of the definitions that were offered were:

\par\quad
\name{Widom}: `a group of atoms that [repeats and] fills space'.
\par\quad
\name{Rabson}: `a geometric assemblage that maximizes coordination number'.
\par\quad
\name{Henley}: `a group of atoms which always stay together in some ensemble.' 
\par\quad
\name{Pay-Gomez} amended this to `the largest group of atoms which ...' 
\par\quad 
[`Stay together' was meant
here in a geometrical or statistical sense, not a bonding sense.]

\noindent
\name{Henley's} point was that the ensemble is implicit whenever
we talk about clusters. That is, a `cluster' must appear in
more than one environment, or in more than one structure.
[\name{Pay-Gomez} commented that finding a cluster on the basis of
just (say) the 1/1 approximant of the i-CdTb \quasicrystal is
like estimating the slope of a curve when you have just one point!]
When we are concerned with energetics, this ensemble would be
the set of all low-energy structures; when we are concerned with
cleavage, the ensemble would be the cleavage surfaces after
many trials with equivalent conditions.

  \name{Trebin}:  The definition of `cluster'
must incorporate the following points (i) these repeating units must
{\it cover} most of the atoms; (ii) The cluster centers provide a
{\it coarse-grained} description of the system's physics;
(iii) the clusters are often easier to {\it observe}, e.g. in
electron diffraction.
[This is elaborated later in subsection~\ref{sec:cluster-Trebin}.]

\name{Naumis} concurred with the viewpoint of {\it coarse graining} [as being important
in the cluster idea], since usually there is a 
hierarchy of  forces, divided into  intra- and inter-cluster forces.
In agreement, \name{Theis} said the reason we need `clusters', is to
implement a new `coarsened' picture of the system
that embodies the physical properties.

\name{Grin} believed the term `cluster' ought to have a meaning similar
to its chemical definition. That is, the interaction within a `cluster'
differs from [is stronger than] that outside the `cluster'.  Any attempt
to define structures geometrically has an arbitrariness.
[But until the higher relative stability of the \quasicrystal
(QC) clusters with respect  
to their environment can be clearly confirmed, it would be reasonable  
to state that the term `QC cluster' is used in the sense of the  
system of nested polyhedra.]

\name{Kreiner} stated that, 
as a chemist working in the field of \quasicrystals, he frankly 
does not use `cluster'  in its original, chemical sense, which
would require us to first solve the problem of bonding [for the
material in question].  A working definition instead would be
`a geometrical entity which can be used for prediction.'

\name{Zijlstra}
expressed support for \name{Grin's} view:
if the standard meaning of `cluster' in
chemistry is `stable structural unit', he strongly feels that we
should adhere to this definition.
Too many words are being used differently by
\quasicrystal scientists than by all other physicists. 
To reduce the risk of confusion, 
let us call these objects, for example, `\quasicrystal clusters';
just as people know `herbal tea' is not really tea, we can 
assume they will infer that `\quasicrystal clusters' are not necessarily 
real clusters (in the sense of stable structural units).
[\name{Henley} asks: what about the other usages of `cluster' 
e.g. the `Mayer Cluster Expansion' in statistical mechanics?]
It would be great if we distinguished these different meanings by qualifying 
them as, e.g., `covering cluster', `stable cluster', `\quasicrystal cluster' 
(for the polyhedral shells used in structural models), etc.

\name{Cahn}: A definition must be short and it must be
easy to test whether it applies in any case.
The second criterion leads him to favour
a geometrical rather than an energetic definition.
A good definition
should be fruitful too: e.g. if we find clusters frequently, we can
infer or guess them in another material.

Clusters are clearly well-defined in
a vapour phase.  And many of these ideas apply to molecular
solids, e.g. benzene in a benzene crystal.  But metallic
bonding has a longer range than chemical bonding.
Furthermore, in a truly quasiperiodic structure
[e.g. the Penrose tiling with inflation symmetry],
 we have clusters at all scales.  We need a {\it useful}
definition of `cluster'!  \name{Cahn} asks, must there be a clean separation
of a cluster from the matrix which surrounds it?

\name{Ben-Abraham:}
 An isolated cluster of atoms, such as produced for instance by
evaporation, is well defined geometrically, chemically, and physically.
What we are dealing with in the context of \quasicrystals, 
complex (inter)metallic alloys,~\footnote{
\name{Ben-Abraham} suggests the abbreviation CIMA 
(in place of the now accepted CMA) which 
(a) is more adequate; (b) can be pronounced as a word;
(c) means `the top, summit' in a variety of Romance languages (!).}
and more generally aperiodic condensed matter, is quite another thing.
In this context, `cluster' is a flexible conceptual tool, mostly a shorthand
expression.  We usually designate as a cluster an arrangement of
coordination polyhedra, perhaps with some peripheral additions;
for the ones with accepted names (`Bergman' or `Mackay' clusters)
this suffices to specify what is meant by cluster.
Such an `cluster' may, in each case, be well defined {\it geometrically} but 
has hardly any precise {\it chemical} and/or {\it physical} meaning, 
and perhaps it is not possible to assign one.

Hence, it is impossible and unnecessary to try to establish a formal
definition of a cluster.
When dealing with an abstract geometrical structure (a packing, a tiling, a
covering, a point set) it is preferable, in \name{Ben-Abraham's}
humble opinion, to refrain from using the term `cluster' and, instead, 
to talk about `a patch'.

\name{Roth} considered it important to mention that
the familiar clusters appear nearly always to be spherical or
[as in decagonals] cylinders of infinite extent in one direction.
They are never elongated like an ellipsoid, although that would not 
be unnatural in certain of the tilings.
[That is, perhaps people have an unconscious {\it a priori} bias towards
particular kinds of clusers, and have not necessarily taken alternatives 
into consideration.]
Or should the definition of `cluster' explicitly restrict the shape?
\name{Mukhopadhyay} wondered, should we impose a lower bound on
the size of a `cluster'; does it need to have multiple shells?
\name{Henley} and \name{Mihalkovi\v{c}} noted that, since the
interactions responsible for stabilisation are long-ranged in
metals -- and oscillating,  as seen in pair-potential modeling --
it might even make sense to define a `cluster' in the shape of an 
annulus or decagonal ring, without including the interior atoms
[Those could be in varying configurations, as in `pseudo-Mackay'
clusters, or the i-CdYb cluster.]

\name{L\"uck} 
agrees that the term 
`cluster' has been applied with a variety of meanings to 
different properties of matter. 
For instance, in Kikuchi's `cluster variation method'~\cite{CVM79},
`cluster' means a geometrical construct to compute the configuration entropy 
of a given material. 
Again, the term `clustering' is used to describe the atomic arrangement 
of an alloy when neighboring atoms are observed to be identical species 
more frequently than in a random arrangement; it describes the 
opposite of chemical short range order.

These two examples show that there is no general, unique understanding of 
the term `cluster'; in both cases, 
nothing is said about binding energies within or out of the 
assumed cluster (as was proposed to be a condition for
`cluster' by \name{Grin}). 
\name{L\"uck},  in any article mentioning a `cluster',
would define or explain this term in the introduction to avoid confusion, 
just as any unfamiliar abbreviation or acronym should be defined.

\name{Mukhopadhyay} believes we all understand `cluster' 
to mean some grouping of atoms possessing short- or
medium- range order, and having some properties different from rest of the 
atoms.  This can also be considered as a motif or building block in the solid 
structures. For example, in the $T$-Mg$_{32}$(Al,Zn)$_{49}$ and 
$\alpha$-AlMnSi approximants, 
the motifs are the Bergman and Mackay clusters, respectively.
Those clusters are basically non-overlapping and non-touching motifs 
decorating the unit cell.  
In respective crystalline, icosahedral, and glassy phases these 
clusters are arranged periodically, quasiperiodically and randomly. Therefore
the clusters can be thought of as a group of atoms acting as non-overlapping 
motif; they govern the properties that depend on the interplay of
short-range (or medium-range) order with the crystal chemistry.
structures. 
[In the $T$-Mg$_{32}$(Al,Zn)$_{49}$ 
case, the Bergman  cluster notion was productive, 
as it inspired the search for the non-crystalline allomorph.]

\subsection{Theorists' definitions of clusters connected to 
binding or physical properties}
\label{sec:cluster-Trebin}

As \name{Trebin} understands them,
clusters are identical groups of atoms which in
most cases can be inscribed into spheres or circles, in rare cases
into other compact sets. These spheres can also overlap, and there may
be more than one type of cluster. 
The clusters thus usually comprise more than 80\% of the atoms. 
A coarse grained description of
complex alloy structures is then given by the positions of the cluster
centers and the tilings formed by them. 

But clusters must have also a minimum stability, this is one way to 
interpret `stay together' in the definition proposed by Henley. 
Hence they are also {\it physical} entities with
adhesiveness in a certain energy range. This definition is suitable
e.g. for the pseudo-Mackay clusters of i-AlMnPd. Numerical simulations
of crack propagation and comparison with STM investigations strongly
indicate that dynamical cracks pass around these 
clusters~\cite{roesch05}.
The clusters are present in the icosahedral quasicrystalline forms and the
$\xi$-approximants of Al-Mn-Pd and thus appear to rearrange themselves as
whole entities during mutual phase transformations. They also stay
together when under plastic deformation partial dislocations and the
associated metadislocations are formed and start moving~\cite{engel05}.
Such a cluster motion need not necessarily happen by simultaneous motion 
of all cluster atoms. Indeed, in some instances like Al-Cu-Co motions of
a few atoms mimic the motion of entire clusters~\cite{zeger96}.
Nevertheless \name{Trebin} would call such clusters physical entities. 
So far clusters are not comparable with molecules although these, too, 
do dissolve and recombine in dynamical equilibrium.

\name{G\"ahler} asks, where do draw the boundary of a cluster, 
and just when can we call it a cluster at all?
Obviously, if a certain arrangement of atoms occurs all over the
place, this arrangement must somehow be preferred, energetically
or kinetically. 
But it is highly problematic to demand that the cluster and the rest 
of the system should be distinguished by the degree of binding. 
Such a notion would make sense for covalently bound entities, 
but those are certainly not typical in metals.
In metals -- including complex intermetallics such as \quasicrystals --
the binding notion
simply does not have a sufficiently localised meaning, as
is evident if we try to 
draw a boundary between the preferred arrangement
and the rest of the system: 
the interactions' effective range is say 6~\AA{} -- typically much larger 
than an interatomic distance. Having a `cluster' embedded in the
rest of the system, it is thus not well-defined which [atomic]
interactions are intra-cluster, and which are between the
cluster and the rest of the system; and
whether a particular `cluster' of atoms is preferred must
depend also on the environment in which that `cluster' is embedded. 
If we were to include, in the very definition of `cluster', 
an energetic separation from the environment, 
it would mean that we have to stop using that term!

Another point is that clusters often overlap, covering a large
portion of the \quasicrystal. Where would one draw the boundary 
[to apply the bonding criterion].
If all of these were to be tightly bound clusters, the \quasicrystal would 
become a single large molecule, with some extra glue atoms in between. 
\name{G\"ahler} doesn't think this is the correct picture. 

\name{Zijlstra} acknowledges that a problem with 
the definition of `stable clusters' 
as having stronger bonds within the unit than between units, is
that in \quasicrystals we often see overlapping (interpenetrating) clusters, 
sometimes covering almost all the atoms.  And for [assessing the stability 
of] clusters that appear in the bulk, interactions with the 
mainly Al surrounding atoms, at least, must be taken into account.
So, the bond strengths
should be the target of theoretical and experimental investigations,
to determine which building blocks are more stable than others  
(so they gain more than their present, mainly geometrical, role).

\subsection{On clusters and ico \quasicrystals described by cut-and-project}
\label{sec:cluster-6D}

\name{de Boissieu} made the following comments on the application
of clusters in \quasicrystals (QC). 

First, for a cluster description to be meaningful it must consist of both a
geometrical network (topology) and of a corresponding chemical decoration.
Second, for a structure as complex as a \quasicrystal, one should use all 
possible descriptions,  and consider which is appropriate for 
a given problem. 

(i) If any physical property is to be deduced from a cluster
description of the QC, its chemical decoration (chemical order or
disorder) will be crucial. Some answer to this question of chemical
decoration can certainly be obtained from the most recent QC structural
refinements.

(ii). Besides the topology of the cluster network, there are also
tiling descriptions of the structure. The restriction to be a tiling
is certainly useful for modeling the atomic structure, but is it really
valid for physical properties calculation?  Are there any strong
experimental evidences for an underlying tiling? [Note that most tilings
constructed from HRTEM images end up with quite a lot of disorder.]

(iii).  Surface studies of QCs are a good example where 
the cluster description has led to little
understanding of the surface structure. A description of the QC structure
in term of dense planes (as already evidenced in the early models of
i-AlLiCu) has been a much more fruitful approach.

(iv). Among the various `angles' from which \quasicrystals have been 
modeled, the hierarchic [inflation] nature of \quasicrystals has been 
little explored for physical properties (except by 
C.~Janot, see e.g.~\cite{janot94}).
Yet this is a striking feature of the diffraction pattern of QCs, 
and \name{de Boissieu} suggests it should be better explored from 
the viewpoint of understanding the physical properties.

\name{Papadopoulos} pointed out, subsequent to the discussion,
that in ~\cite{zorka99} and subsequent
papers, she and collaborators demonstrated that [assuming no
reconstruction, and quasiperiodic models of the accepted type] 
{\it any} surface on i-AlPdMn [or i-AlCuFe] must cut through {\it some} Bergman,
or Mackay, clusters.  This shows that clusters are `not stable',
in the energetic sense.

\name{Barbier} critiqued the entire notion of cluster descriptions.
Within the aperiodic structure of \quasicrystals (QCs), some local atom
configurations can be identified:  for instance, Bergman or Mackay-like 
clusters for icosahedral \quasicrystal (i-QC) structures [of the Al-TM type].
Are these groups of atoms relevant in the QC structure? 
Certainly, the symmetry of QCs causes some local configurations to be 
repeated in the structure. 
He noted the following points:
\begin{itemize}
\item [(i)]
One may extract from the i-QC structure the aperiodic distribution
of Bergman clusters. 
\item [(ii)]
On the other hand, one could describe the very same atom distribution 
with Mackay clusters instead.
\item [(iii)]
Furthermore, excess atoms (`glue atoms'), not part of the clusters, are
present in the structure.
\end{itemize}
Thus, [from (i) and (ii)] the cluster description is not unique,
and [from (iii)] does not catch the whole
structure.
[\name{Barbier} instead 
advocated a structure description completely founded on the
hyperspace cut approach, see section~\ref{sec:cluster-surface-Barbier}.]

\subsection{Evidence for clusters from electron density measurements}
\label{sec:cluster-Kimura}

In this subsection, \name{Kimura} addresses the questions
posed in \name{Steurer's} discussion paper~\cite{steurer-qdisc},
based on two of his papers
(Refs.~\cite{kimura-JAP03}  and \cite{kimura-PRB03}),
containing experimental data from the
1/1 approximant $\alpha$-AlReSi of i-AlPdRe, in which
Mackay Icosahedron (MI) clusters are present.

\begin{itemize}
\item[Question (i)]
How is the distribution of chemical bonds (length, strength, type, anisotropy)
between atoms in a geometrical cluster found from structure analysis? 
Are the strongest bonds between atoms of a shell or between atoms 
of different shells? 
Is there a difference between the bonding of different shells 
(decrease in bond strength from the inner to the outer shells, 
not every shell consists of atoms in bonding distance to each other)? 
How does the network of strong bonds look like for concrete examples?
\end{itemize}

The electron density distribution in the $\alpha$-AlReSi 1/1-approximant
crystal was measured using synchrotron radiation combined with a 
fit by the maximum entropy method~\cite{kimura-PRB03}.
The criterion to define bond-strength was the electron density at the
bond midpoint.
There is a wide distribution of the bond-strengths in $\alpha$-AlReSi, 
ranging from strong covalent bonds (near to those in Si) to weak metallic bonds 
(near to those in fcc Al), which is considered to be one of the features 
of a structurally complex material.

\begin{itemize}
\item[Question (ii)]
Is it possible to identify clusters clearly separated from the embedding 
matrix?  What is the size of these clusters, how thick are the matrix parts 
between clusters?  
What is the difference between cluster and matrix (chemical bonding)?
\end{itemize}

According to the above estimation, the intra-Mackay cluster bonds
are stronger than the inter-Mackay cluster bonds, on average. The
$\alpha$-AlReSi and also i-AlPdRe \quasicrystal are considered to be
intermediate states between metals, covalent bonding networks
(as in semiconductors) and molecular solids.

\begin{itemize}
\item[Question (iii)]
Is it possible to model in a first approximation the physical properties 
(electronic, dynamic) of \quasicrystals in terms of clusters embedded 
in a matrix?
\end{itemize}
 
     The thermoelectric figure of merit $Z$ and the effective mass $m_*$ of
i-AlPdRe \quasicrystals can be increased by strengthening the intra- and
weakening the inter-cluster bonds. According to this scenario, $Z$ was
improved by substitution of Ru for Re~\cite{kimura-JAP03}.

\subsection{Evidence from surface experiments
regarding the stability of clusters in \quasicrystals}
\label{sec:cluster-McGrath}



Here \name{McGrath} and \name{Pat A. Thiel} review the
experimental evidence for enhanced stability of clusters from 
cleavage experiments under ultra-high vacuum conditions. These 
include:
\begin{enumerate}
\item [(i)]
The experiments of Ebert, Urban and co-workers on cleavage in
ultra-high vacuum conditions of the 2-fold and 5-fold i-AlPdMn
surfaces~\cite{Ebert96,Ebert98,Ebert99};
\item [(ii)]
Similar experiments by Ebert and co-workers on the 10-fold surface of
decagonal AlNiCo~\cite{Ebert03};
\item [(iii)]
experiments by Cecco and co-workers on the cleaved 10- and 2-fold
surfaces of d-AlNiCo and the pseudo-10-fold surface of the
$\xi^{\prime}$-AlPdMn approximant \cite{Cecco04}.
\end{enumerate}
Scanning tunneling microscopy (STM) measurements of all of these 
surfaces formed
by cleavage showed the presence of approximately nanometre size protrusions.
These surface experiments are practically the only experimental pieces
of evidence for the physical reality of clusters in \quasicrystals.

Considering first the cleavage of AlPdMn surfaces, the roughly 1 nm 
protrusions seen on the 10- and
2-fold surfaces of i-AlPdMn were linked by Ebert {\it et al.} to the
enhanced structural stability of Mackay-type clusters. For the 2-fold surface,
autocorrelation patterns of the images showed some order in the arrangements;
none was found for the 5-fold surface. On the other hand, Cecco 
\emph{et al.} \cite{Cecco04}
found that  the cleaved pseudo-10-fold surface of
$\xi^{\prime}$-AlPdMn showed protrusions the order of 4--8 nm in size, with no
evident preferential orientation and no periodicity.
As the unit cell parameters of $\xi^{\prime}$-AlPdMn are $a$=2.3541 nm
and $c$=1.2339 nm, the authors conclude that there is no evident correlation
to the underlying structure, which contains partial Mackay
icosahedra~\cite{Boudard96-PMA}.

Turning to the 10-fold surface of decagonal AlNiCo, the 1--2 nm size 
protrusions seen on this surface
were interpreted differently by the two groups that observed them:
Ebert {\it et al}~\cite{Ebert03} saw them as evidence for the 
existence of columnar clusters in $d$-AlNiCo. On the other hand, 
Cecco {\it et al}~\cite{Cecco04} could find no correlation
between the protrusions seen in their images from the 10-fold surface and the
underlying structure.

What are the corresponding results when {\it simple} metal crystals 
are fractured?
Although the literature does not appear to be extensive, STM 
measurements from cleaved
Bi(0001)~\cite{Edelman96} and Sb(0001)~\cite{Stegemann04} show these surfaces
as being atomically flat over large areas.

To summarise, therefore, all evidence is that cleaved \quasicrystal
surfaces are rough on the nanometer scale (also found in the approximant
$\xi^{\prime}$-AlPdMn), but there is disagreement as to whether this
indicates enhanced stability of pseudo-spherical (or columnar) 
clusters. The roughness might also be explained in terms of the 
non-existence of  natural cleavage planes in these complex
alloys: in the bulk structure of i-AlPdMn and i-AlCuFe the 
largest `gaps' perpendicular to the 5-fold direction (evident in 
Yamamoto's structure model) are of the order
of 1 \AA{} thick~\cite{Sharma04}, which is just half of the (0001) 
interplanar spacing (about 2 \AA{}) that occurs in Bi or Sb 
\cite{Edelman96,Stegemann04}.

The limited data available need to be augmented through
further experiments. These might include fracture of other refractory metal
surfaces along high index planes, and fracture of other complex metallic alloy
giant unit cell structures. 
As to repeating the fracture experiments on \quasicrystals,
it should be kept in mind that  these samples are very precious
[and are destroyed by the experiment].

As a final remark, \name{McGrath} and \name{Thiel}
would echo the comments in C.~L.~Henley's
discussion paper~\cite{henley-qdisc},
that regardless of whether clusters have a physical reality
`they are inescapable as a framework to organise our understanding
of a structure'. In STM measurements of the icosahedral five-fold surfaces,
truncated pseudo-Mackay and pseudo-Bergman clusters appear as identifiable
characteristic motifs, which also turn out to be important in adsorption
and nucleation processes. Clusters, whether stable or not, can provide a
common language between the experimental bulk, experimental surface and
theoretical communities.

\subsection{Clusters: surface science and 6D-cuts}
\label{sec:cluster-surface-Barbier}

This subsection consists of a critique by 
\name{Barbier}, wherein surface studies furnish 
additional arguments about the role of clusters
[see his comments in subsection~\ref{sec:cluster-6D}, above].

Let us first recall the height correlation function 
   \begin{equation}
        G({\bf r}) \equiv \langle [h({\bf r})-h(0)]^2\rangle
   \label{eq:htcorr}
   \end{equation}
used to characterise the roughness of any surface.
On terraces of a well equilibrated [facet] surface,
$G({\bf r})$ saturates at large distances, which means the terraces
are {\it flat}. 
On the other hand, 
a thermally {\it rough} surface (as it is generally the case for low
index vicinal surfaces of usual crystals) has a long-range height correlation
divergent  as $G({\bf r})\sim  \ln(r)$ (and no faster: that is a
universal property of {\it equilibrium} 2d interfaces).

\subsubsec{Well equilibrated \quasicrystal surfaces}

For the 2- 3- and 5-fold surfaces at thermal equibrium, 
STM observations show wide 
terraces; the helium diffraction experiment shows that this is valid 
for the whole surface. 
Analysis of the surface structure shows that the terrace planes cut
the Bergman and Mackay clusters~\cite{barbier02}.

As in the bulk, on extended flat terraces of 
icosahedral \quasicrystals (i-QCs), 
e.g. i-AlPdMn~\cite{barbier02} or i-AlCuFe~\cite{cai02},
well-defined patterns are seen which could define surface clusters:
filled (or empty) flower-like patterns, 10-fold rings of pentagons.
These are useful as a guide
for the eye to perceive the QC symmetry (pentagonal structures,
aperiodicity, rows following Fibonacci sequences),
yet one cannot describe the surface structure solely on this basis. 
A much more fruitful approach is the 6D description of i-QC structures. 
Given one unique periodic (within the 6D space) unit cell, all possible
configurations within the real 3D space can be obtained allowing a complete
description of both terrace structures and step height distributions.

\subsubsec{Cleavage}

It has been proposed that fracture surfaces found in
experiments on \quasicrystals~\cite{Ebert96,Ebert03}
exhibit at the lowest scale a cluster-like pattern. 
In fact, experiments on {\it all} materials have found
a universal behaviour of the roughness of relaxed fracture
surfaces~\cite{bouchaud97}.
All fracture surfaces exhibit a self similar
roughness within the range $\xi_c$, the elementary element size, up to 
$R_c$, the limit at which linear continuous mechanics applies. 
Within this range, the
height correlation of the self-affine morphology follows a power law:
$\sqrt{G}(r) \propto r^\alpha$ with the universal exponent $\alpha = 0.8$. 
[This is a greater roughness as compared to the thermally rough surface.]
Furthermore, this exponent $\alpha$ is found to be independent of the 
material structure~\cite{bouchaud97}: it is the same in 
glass, intermetallic alloys, composite materials like concrete... 
even in  single crystals~\cite{deegan03}.

Up to now, measurement of $\xi_c$ has not been performed for QCs.
Only a scale analysis of the roughness of cleavage surfaces over 
several decades of length scales would allow one to conclude in favour 
of a possible relevant cluster morphology in cleavage surfaces.
[Deviation from the universal law at low length scales could give 
indication of some elementary size -- atom, cluster, microcrystallite.]

\subsubsec{Hyperspace description}
\label{sec:cluster-Barbier-hyperspace}

Thus, leaving the various attempts of a surface cluster description, surface
studies better contribute to unambiguously define the unit cell within the
associated hyperspace~\cite{barbier02,cai02}.
Only this unique and unambiguous (even if not
yet definitely fixed) unit cell allows to capture 
all the richness of the QC structure.
The origin of the stability of icosahedral QC structures necessitates a
better understanding of what is allowed in the decoration of the 
6D unit cell (the geometry being fixed by the symmetry properties). 
The energetic balance that would be deduced within the hyper space 
description is still lacking. 
Devising a theoretical framework to calculate energies
within a 6D framework
is the only way to gain a deep understanding of QC
structures.

Reducing the QC description to the distribution of some clusters
would miss the fundamental [defining] property of QC: their aperiodicity 
as a consequence of the symmetry [of their atomic structure], which can
only be completely represented in the hyper space geometry.
Clusters of {\it various} positions and sizes can be built from this 
well-defined [hyperspace] unit cell. 
Though they are useful for immediate comparison of structures with
different local configurations (thus, for revealing whether the 
6-d unit cell has different contents) they cannot catch
the essence of the QC structures. Within a QC structure clusters
cannot be defined in a unique way; whereas, as for usual crystals, unique
definitions of both symmetry and hyperspace unit cell satisfy the
crystallographer. 

\subsubsec{Conclusion}

To summarise the main points:

(i) Though a cluster view may be useful for a quick comparison of various 
surface (or bulk?) structures of QC, it is useless (or misleading ...) 
in a description of the whole structure. 

(ii) Experimental proof of the existence of clusters by cleavage experiments
is far from proven,  so long as a scale analysis of the surface height
correlation function is lacking. 

(iii) Assuming the unit cell within the hyperspace description is a
reality, the following definition of QC clusters would be
implied:

\begin{itemize}
\item[]
{\it `QC Clusters' are various [frequently repeated] assemblies of 
[several] atoms that can be generated and are distributed according to the associated 
Bravais hyperspace lattice decorated by one single unique unit cell.}
\end{itemize}

A cluster approach to describe QC structures will remain
necessarily ambiguous. [The better way to use clusters is
in the context of the rest of the \quasicrystal -- which  is
specified in 6D -- and they should be defined so as to have
a simple relation to the 6D cell.]


\section{Phason fluctuations}

{\it This discussion was led by \name{Marc de Boissieu}.}

\subsection{Definition of `phason'?}

This term probably occasions even more confusion than `cluster'
in the field of \quasicrystals.
\name{Widom} began with a question:
what is \name{de Boissieu's}
meaning for `phason hopping'? If an atom
hops back and forth [between the same two places],
would you call it a `phason'?
(See \name{de Boissieu's} discussion paper~\cite{deboissieu-qdisc}.)

\subsubsec{A critique of the `-on' in `phason'}

\name{Ben-Abraham}
emphasized that strictly speaking, `X-on' is a quantum of excitation X.
Thus, a `phonon' is one quantum of a quantized lattice vibration mode.
That is manifestly evident in the occupation number representation.
However, it is commonplace to refer, by metonymy, to the lattice vibration
mode itself as a `a phonon'.  That is not quite correct, but acceptable and
generally also accepted.

In the context of aperiodic crystals we deal with excitations called
`phasons'.  In incommensurate structures these are precisely defined and well
named.  In more general cases, perhaps less so.  Yet they clearly refer to
the extra degrees of freedom present in aperiodic crystals.
Flips are, of course, their elementary manifestations.
Nevertheless, it is grossly misleading to call the single flips `phasons'.
(\name{L\"uck} concurred with 
\name{Ben-Abraham's} comment that the term `phason' is 
misleading due to the ending `-on'.)

In the early days, \name{Baake}, \name{Ben-Abraham},
and the T\"ubingen group used to call
these flips `simpleton flips' as they occurred in the simplest hexagonal
vertex (= local environment);
we now know that they occur in more complex environments, too.
To conclude, everyone knows what is meant by a flip in a \quasicrystal.
So, \name{Ben-Abraham} proposes, let us call it just `a flip'.
It does not need the {\it epitheton ornans} (superfluous adjective)
of `phason' in front of it.

\name{Henley} offered a polemic about nomenclature that is 
introduced carelessly.  The late Per Bak first used
the term `phason' for \quasicrystals~\cite{bak85}
but afterwards, at least once he 
claimed he meant to write `phase',  e.g. `phase stiffness'.
Thus, for us, `phason' is properly only an adjective, e.g. 
`phason mode' or `phason coordinate', and is similar
in meaning to `perp-space' or `complementary-space'.
Yet the {\it name} ending in `-on' suggested a discrete, 
countable object, so some people attached it to the matching-rule 
violations or to dynamic tile flips,
which implement phason modes in discrete models.

\subsubsec{Phason defects: point, line, or wall?}


\name{L\"uck} commented on the
the geometry and observations of 
{\it phasonic defects} as described in 1988 and 
subsequently~\cite{lueck88,lueck93+,lueck94}.
(In this viewpoint, a phason-free \quasicrystal would result from an
flat cut at the correct slope through the appropriate hyperspace
crystal.  A `phason' refers to an elementary violation of a local
rule, such that, furthermore, a closeness condition in complementary 
space is fulfilled.~\cite{lueck94}.)

In 3D, a phasonic defect is a 
{\it line} defect forming a closed loop or terminating at the surface or at a 
dislocation line; only in 2D is it a {\it point} defect described by a jog in 
the so-called `Conway worm'. Movement of phasonic defects requires phason 
flips, usually realized by jumps of atoms (or groups of atoms), 
as described in the discussion paper by 
M. de Boissieu~\cite{deboissieu-qdisc}.

The remarks in this discussion by \name{Trebin} and 
detailed investigations of M. Feuerbacher {\it et al}~\cite{feuerbacher}
mentioned the interaction of dislocations and phasonic defects as line
defects.
Details concerning the healing of phasonic defects formed 
in plastic deformation were discussed in a collaboration by \name{L\"uck}
and David Warrington~\cite{warrington01}.
The formation of `phason walls' is definitely 
compatible with the required property of a line defect.
A high concentration of phasonic defects may 
destroy the line character of phasons due to the high density. 
However, there is no reason that a phason is regarded as a point defect.

\subsection{Theorists' viewpoints on phason elasticity and dynamics}

\subsubsec{Hydrodynamics: need to model short wavelengths}
\par

\name{Trebin}
distinguished `phason flips', which correspond to jumps of one atom or a
localized group of atoms between split positions, and 
`phason excitations', which consist of correlated phason flips over 
large distances that together make the macroscopic shift.

Long-wavelength phason excitations are well described by
the hydrodynamical equations established by 
Lubensky {\it et al}~\cite{lubensky85} and others.
Some phenomena like phason-induced diffuse scattering and diffusion 
of phason fluctuations ~\cite{francoual03}
[seen in X-ray speckle imaging of diffuse wings], and 
also `phason walls'~\cite{phason-wall}
[created by moving dislocations],
have been studied experimentally.
The decay time constant in 
the mechanical experiments is similar to that
for fluctuations in the diffuse scattering 
experiments [on similar length scales].
Thus, the existence of the phason degree of freedom in \quasicrystals is 
well established. 

What is missing up to date are studies of phason dynamics
beyond the hydrodynamical limit, i.e. phason excitations of shorter
wavelengths, phason damping mechanisms and the freezing of phason
kinetics or `phason pinning'. For the study of these phenomena,
discrete models are required~\cite{trebin05}
and (\name{Trebin} speculated) 
concepts from the theories of the glass transition.

\name{Rivier} commented that years ago, there was an interesting theory
by Kalugin and Katz, about how the brittle-ductile transition
in \quasicrystals was due to a percolation of phason flips.
Gratias pointed out that experiments contradicted that theory:
they didn't actually show the catastrophic change of atomic
diffusion at high temperatures, which it predicted.

\subsubsec{Hydrodynamics: need to measure long wavelengths}
\par

In this subsection, \name{Ishii} reviews
the different behaviours, or even different physics, that are
expected for
phasons in the short and long wavelength regimes,
stressing the {\it dynamic} aspects of phasons in these different regimes.
An ideal quasiperiodic structure is obtained by a flat cut
${\bf h}_\perp=$constant [here ${\bf h}_\perp$ is perp-space coordinate] 
of higher-dimensional lattice but
a cut for real samples is inevitably corrugated somewhat. Such
corrugation or fluctuation in perp space is induced, for example,
due to uncontrollable experimental conditions such as inhomogeneities
in temperature and compositions at the growth front.

According to the hydrodynamic theory, spatial Fourier components of 
the phason fluctuation with wavevector ${\bf q}$ relax with a time
constant proportional to $1/|{\bf q}|^2$. The short-wavelength
(large $|{\bf q}|$) component is thus rapidly relaxed, whereas the
long-wavelength one cannot be relaxed completely. Such unrelaxed
fluctuation is a kind of disorder and yields diffuse scatterings
around Bragg peaks. Depending on the phason elastic constants
(so-called $K_1$, $K_2$ and $K_3$ for the icosahedral case, 
where $K_3$ is the phonon-phason coupling)
one particular mode of the fluctuation has a larger time
constant than others and eventually dominates the phason disorder.
This is a scenario proposed by Widom and Ishii~\cite{widom-ishii91}
 and is exactly what de Boissieu observed in i-AlPdMn~\cite{deboissieu95}.

In this case of the long-wavelength
phason fluctuation, individual phason flips, which are local
rearrangements of underlying tiles or atomic jumps, correlate at
a long distance because 
the same Fourier mode is responsible for  them.
This correlation is an important consequence of the
quasiperiodic order in the system.

Edagawa~\cite{edagawa00}
observed local  phason flips in high-resolution electron microscopy.
It would be very nice if he also observed that
long-wavelength components of the phasons remain 
unrelaxed in the sample after long-time measuremen, but
apparently this has not yet been tested.
For the jumps observed by \name{Edagawa} 
(also Coddens's time-of-flight measurements~\cite{coddens99}),
\name{Ishii} believes
the short-wavelength component of the fluctuation and its fast dynamics
play primary roles. The hydrodynamic description is not
appropriate enough for such flips localized in real space.
In situ observation of the relaxation process, if possible,
would be interesting and important for further understanding of
the phason degrees of freedom in \quasicrystals.

\subsubsec{Phason elasticity from other models?}
\label{sec:phason-widom}

\name{Widom} called attention to the issue  of 
the derivation of phason elasticity from the various stabilisation 
models.  To date the square-gradient elasticity has been derived 
only for random tiling models, and appears to be valid for 
matching rule models in their high temperature `random tiling'
phases.  He is not aware of the status of elasticity properties 
for models based on either pseudogaps or on cluster density 
maximisation;  he would encourage people advancing such models to 
determine their elastic properties.  At the same time, \name{Widom}
would appreciate a survey of the experimental support for 
square-gradient elasticity and an assessment of whether it 
should be considered as experimentally proven.

\subsection{Experiments on phason dynamics}

In brief, 
it appears that a large phason mode occurs, which is a combination
of many elementary flips;  But it is not sure this is part
of a long-wavelength phason mode.

In this subsection, ~\name{Edagawa} addresses
\name{Widom's} question, 
-- is there agreement as to what Edagawa observed in the [video] 
HRTEM experiments~\cite{edagawa00}, 
that claimed to see phason hopping? -- 
and some related issues.
Note that in this work, we can construct a tiling structure 
from the image without ambiguity. The tile vertices can be lifted up to the 
high-dimensional lattice, from which we can define the spatially varying 
phason displacement ${\bf h}_\perp({\bf r})$
(a smoothed function obtained from the staircase 
structure in the high-dimensional space).

\smallskip
\noindent{\it 
Question 1. What is observed in Edagawa's experiments?}

It is a going-back-and-forth phason flip in a tiling
with edge length 2 nm. Of course it is 
not a long-range phason fluctuation itself but a localized phason flip.

\smallskip
\noindent{\it 
Question 2. Many theoreticians think the term 'phason' should be used 
only for a long-range mode.
Therefore, should such a localized flip be 
called just 'flip' or 'flippon'?.}

\name{Edagawa} does not think so. 
The flip observed is not just 
a flip but a flip leading to the change in the local value of 
${\bf h}_\perp ({\bf r})$ [constructed as explained above].
Hence, such a flip is properly called `phason flip'.

\smallskip
\noindent{\it 
Question 3. 
How can one relate such a local phason flip to the long-range phason 
mode?}

First of all, such a long-range phason fluctuation must consist of a 
combination of local phason flips. 
It is not sure whether the particular phason flip 
observed in \cite{edagawa00}
is part of a long-range phason mode fluctuation. 
We attempted to deduce spatial and temporal correlations from 
our movie, but we could obtain no clear result due to limited area-size and 
time-duration in the observation.  However, we can safely say at least that 
what we oberved is a local fluctuation of $h({\bf r})$, 
as Prof.~\name{Ishii} pointed out.

\smallskip
\noindent{\it 
Question 4. 
The time scale in the phason flip we observed is by many orders longer 
than those observed using neutron and M\"ossbauer by Coddens {\it et al}. 
Why?}

What they observed is an elemental phason flip involving only a single atom 
(or a few atoms). What we observed 
is a larger-scale phason flip in a 2nm tiling in HRTEM images, 
which are a  projection of the structure (through a thickness of 
maybe 30nm): this flip consists of a combination of many elementary phason 
flips. That seems to explain qualitatively the 
large difference in the time-scale, though no quantitative evaluation has 
been made.

\subsection{Phasons and electronic structure}

\label{sec:phason-naumis}

In this subsection, \name{Naumis} 
has elaborated the connections between notions of average lattice, 
phason disorder, clusters,  and the electronic stability of
\quasicrystals~\cite{naumis05}.
Since we know that phasons exist, and
assuming that \quasicrystals are stabilized
electronically stabilized {\it via} the Hume-Rothery mechanism,
somehow the electronic structure must remain the same when 
phasons are introduced. How is this possible? 

\subsubsec{Phasons and inflation/deflation}
\par

The first question is whether phasons are local jumps or coherent? 
In ~\cite{naumis99} and ~\cite{naumis04},
we showed that even if the [cut] window is
shaken at random in the higher dimensional space, there is a spatial
correlation in the sites where phasons are observed. 
The picture for phasons was found to be similar to that for vacancies: at
small scales, one can describe them as jumps, but at large scales,
conservation laws determine the behavior. 

What is really correlated for phasons  is the {\it probability} 
of having phason jumps~\cite{naumis99}, which is different for
different sites.
The sites with a bigger probability of jumps are in the edges of the
window~\cite{naumis05}; then the probability of phason jumps depends
on the perp-space coordinate in a way related to the deflation rules 
of the \quasicrystal. 
Namely, sites at the center of the cut band, or window, 
are very unlikely to  flip-flop, 
since the window would need a huge fluctuation 
in order to not include such a site.
So, one can define an `effective band width' 
[of these unlikely sites], a reduced window which turns out to be 
just a deflation.
Indeed, it was observed experimentally~\cite{abe03},
that phasons occur in deflated structures.  

\subsubsec{Consequences for electronic structure}
\par

Next we must prove that phasons, in the sites where they are likeliest to
happen, do not compromise the density of electronic states and thus 
do not destroy [destabilize] the \quasicrystal. 
In a Fibonacci chain this was confirmed and in high dimensions it
also seems to be true~\cite{naumis05}:
sites that are far from the center of the cut window do not contribute
so much to the diffraction bright spots,  which determine 
the electronic  gaps~\cite{smith87}.
Another way to express this is that the stable points, corresponding
to the center of the cut window, are
more likely to belong to the `average lattice' points,
which contribute the most to the 
bright diffraction spots~\cite{aragon02}
and thus determine the electronic structure.

\section{Stabilisation of \quasicrystals}

Most participants were aware that there are two general viewpoints
on \quasicrystal stabilisation: energetic stabilisation of an 
ideal quasiperiodic structure, versus random-tiling ensembles
which are presumed to be stabilized (but only at higher temperatures)
by tiling configurational entropy.  This is taken up in 
\name{Henley's} discussion paper~\cite{henley-qdisc}. 
[It also forms one subtext of the phason discussions, since it
is claimed that the hydrodynamic elasticity observed in
diffraction experiments~\cite{deboissieu95,francoual03},
is realizable only in a random-tiling-type state.]
But relatively little of the discussion was devoted directly
to this dichotomy, perhaps because there was little new to say.

\subsection{Mechanisms of stabilisation: miscellaneous remarks}

\subsubsec{Role of diffraction (structural) experiments}

\name{Gratias} stated that ideal models make specific predictions
[about the atomic structure as it appears in 6D].
\name{Widom} replied that the phason Debye-Waller factor that appears
in all refinements [has the physical meaning of random fluctuations
of the phason coordinate].  In effect, this is incorporating
random-tiling effects in your ideal models.
\name{de Boissieu} noted that the
fitted [phason or perp space] Debye-Waller factor,
$DW_{\perp}$, is in agreement with
the measured diffuse scattering. He wondered~\cite{deboissieu-qdisc}
if this parameter is just another way to accomodate
chemical disorders of the atoms? [\name{Steurer} agreed.]
It is important to go measure the diffuse scattering [not just the
Debye-Waller factor], since only this can measure correlations
[and correlations are necessary to characterise the nature
of the disorder].

\name{Trebin}:
urged that we need to look for the temperature dependence of the phason
elastic constants [as a way to distinguish whether its origin
is mainly energetic or entropic]; it would respectively
be decreasing or increasing with temperature.
\name{de Boissieu}
replied that this experiment was already done by his group, and
it shows the phason elastic constant {\it increases}
with temperature.

\subsubsec{Are random tilings inconsistent with line compounds?}
\par

There exist `line compound' \quasicrystals e.g. i-CdYb;
\name{Mukhopadhyay} (with others)
has wondered whether a random tiling  can
possibly have a definite stoichiometry (and thus touches
the applicability of the entropic stabilisation model).
\name{Henley} clarified that 
the random tiling model is still applicable to line compound \quasicrystals,
if the stoichiometry is maintained when tiles are interchanged.

\subsubsec{Hume-Rothery stabilisation}
\par

To \name{Naumis}, 
we must start by considering that \quasicrystals 
are electronically stabilized via the Hume-Rothery
mechanism, i.e., the electronic density of states must develop a pseudogap
at the Fermi energy to lower the kinetic energy per electron. 
(This is similar to the Peierls instability of
one-dimensional metals, a static lattice deformation
that breaks the periodicity.)
This poses interesting questions concerning the interaction
of phasons and the electronic structure, which he elaborated
in that part of the discussion (section~\ref{sec:phason-naumis}).
[\name{Widom} (subsection~\ref{sec:phason-widom}, above)
asked practically the reverse question: what implications
does the pseudogap mechanism have for phason fluctuations?]

\name{Zijlstra} has been confused by the frequent reference to
'Hume-Rothery stabilisation' in \quasicrystals, since he understands
this term to be reserved for $sp$-type materials,  such as i-AlZnMg
[in which plane waves are mixed by potential scattering so as to
split a gap right on the free-electron Fermi surface, thus lowering
the energy since only the bonding state is occupied.]
In icosahedral Al-TM-TM \quasicrystals (and some crystals such as Al$_6$Mn),
it appears instead that the large hybridisation between
the TM $d$-states and the free-electron like $sp$-states
is the main contribution towards stability~\cite{ishii01};
this is visible in the plots of
Trambly de Laissardiere and Mayou~\cite{trambly}
If the $sp$-$d$ hybridisation is turned off by hand 
in the LMTO Hamiltonian~\cite{trambly},
the pseudogap vanishes (to within numerical accuracy).

\subsubsec{Stabilisation and soft matter \quasicrystals}
\par

\name{Dotera} addressed the question,

\begin{itemize}
\item[]
{\it Can \quasicrystals be universal over different length scales? If 
so, do they form with the same origin?}
\end{itemize}

A general belief that, if they exist, all quasicrystalline phases 
are high temperature phases, may be reconsidered when one considers 
the new frontier of soft-matter \quasicrystals. In the `liquid
\quasicrystal' system
reported by Ungar~\cite{ungar-icq9} at this conference, 
the dodecagonal \quasicrystal phase is the {\it low} temperature phase, 
lower than other 3D spherical micellar phases such as 
$P4_2/mnm$ (tetragonal) corresponding to the $\sigma$ phase in the 
Frank-Kasper family. [The hexagonal columnar phase is even lower
in temperature; but that case is a different story since
the spherical micelles made up of dendrons get broken up 
and are reconstructed into cylinders.]
Chemical disorder and tiling entropies discussed in the literature may 
be quite different in soft matter. 
And the length scales grow larger: the tiling edge of the 
$\sigma$ phases are respectively:~\cite{takano05}.
metallic alloy ($\sim$0.5nm), 
chalcogenide ($\sim$2nm), 
dendron ($\sim$10nm), and star block copolymer ($\sim$80nm)

\subsection{Mathematical question of ground-state stabilisation}

\name{Cahn} recalled that some years ago, Charles Radin {\it proved}
that, for `almost all' potentials in a certain mathematical
sense, the ground state is non-periodic.
[\name{Widom} notes that the condition~\cite{radin}
seems to be all $L^1$
normalizable potentials; that is not physically reasonable,
in that the [typical] potential is near zero for most distances but
on a sequence of large and ever-more-widely spaced distances
it has relatively large values.]

\name{Rabson} averred that 
to prove, mathematically, that a \quasicrystal is the ground state of
any realistic system (that is, not of a simplified pair potential designed
to have a quasiperiodic ground state) is likely to be very difficult.
Even for {\it periodic\/} compounds, has any real
structure has been proven (mathematically) to be
the ground state?  Think how long it took just to prove Kepler's 
conjecture about the close packing of spheres.
Aside from the mathematical difficulties, the
temperatures one would like to probe in order to say with some confidence
that a system has a periodic or quasiperiodic ground state are too low
for equilibration; diamond, at room temperature and pressure,
is famously merely metastable.


\name{Rabson} also commented that periodic systems have one advantage 
in securing our faith that they really are periodic:
Bloch's theorem and all that follows from it (band structure and {\it e.g.},
optical and electronic properties) are so successful at describing
and predicting the behaviour of periodic metals, semiconductors, and
insulators that it is difficult to conceive of these materials as other
than (essentially) periodic.  [Of course one can still apply Bloch's theorem
to \quasicrystals considered as the limiting case of a sequence of
approximants, but the predictive value is less.]
What we need are a theoretical framework to replace Bloch's theorem,  
and a set of experimental observations that make the
replacement relevant.

\subsection{Thermodynamic and mechanical experiments}

\subsubsec{Stability of \quasicrystals 
from the viewpoint of phase equilibria}
\par


\name{L\"uck} observes
that we know two main reasons for stability of \quasicrystals. One reason is 
kinetic stabilisation, the other is thermodynamic stabilisation. 
Kinetically stabilized states,
such as the original quasicrystalline state~\cite{shechtman84}
(a melt-quenched Al-Mn alloy)
 are often termed `metastable'. This term does not 
fix the mechanism of stabilisation, 
it only implies that there is an energy barrier to be 
overcome for transition into the stable state \{which was
measured by extensive investigations early in the history 
of the metastable \quasicrystals\}~\cite{kelton85,lueck-review}

For thermodynamic stability of materials,
as pointed out in de Boissieu's discussion paper~\cite{deboissieu-qdisc},
there are two different possibilities.
One is stability as a ground state of matter, at zero Kelvin, 
which can only be determined by heat of formation or enthalpy 
[change] $\Delta H$. In any (theoretical) model several contributions to $\Delta 
H$ should be taken into account. Only pure metals and perfectly ordered 
intermetallics can be the ground state of matter according to the second 
law of thermodynamics; 
so this requires that \quasicrystals in a 
ground state must have a completely determined structure 
with a specified stoichiometry
which rules out the possibility of a random 
tiling model as a ground state.~\footnote{In the case of five-fold symmetry,
the stoichiometry can be expressed by giving each 
element the sum of an integer and a multiple of the golden mean $\tau$.}

The other possibility is stability at ambient and high temperature; 
this is determined by the Gibbs energy  [change]
$\Delta G$, which includes (as is well known) an entropy term.
Contributions to configurational entropy may 
arise from topological and chemical disorder: they
include, in pure metals, vacancies, interstitials and 
vibrations; in solid solutions,
also the entropy of mixing;
and in intermetallics and ordered alloys, 
anti-structural atoms and chemical disorder. 
Quasicrystals have yet another 
contribution to entropy, which is based on phasonic disorder or (for short)
`phasons':  the microscopic realisation
can be topological or chemical or 
a combination of the two types.
At present it seems to be quite certain that \quasicrystals can be 
found in a thermodynamic stable state at high temperature.
Certain quasicrystalline modifications are only 
stable at high temperatures, e.g. the  `basic Ni' i.e. Ni-rich
$d$-AlCoNi stabilized by the entropy of mixing Ni/Co~\cite{lueck00},
also other decagonal or five-fold variants. 
\OMIT{[More examples are described by de Boissieu.]}

\QUERY{Q. DESCRIBED IN HIS CONFERENCE PAPER? OTHER EXAMPLES OF WHAT
- d(AlNiCo) phases?}

\name{L\"uck} considers it a  still open question 
if \quasicrystals can exist as a ground 
state of matter. 
Direct measurement of the heat of formation requires 
investigation of all competing phases of the alloy system, 
e.g. by Sa\^adi {\it et al}~\cite{saadi93}
in the Al-Cu-Fe system.
A conclusion was hard to reach,
since the differences of measured $\Delta H$ values were small;
a second paper~\cite{saadi95} stressed that the 
icosahedral phase can be the ground state, since the values of $\Delta 
H$ {\it versus} concentration form a convex configurational polyhedron.

Experimental attempts to clear up this point
require great care to overcome kinetic difficulties at low temperature, 
e.g. slow diffusion, 
so the equilibration time exceeds a human lifetime. 
There is evidence that many found states, 
especially some of the so-called approximants, are not 
sufficiently in thermodynamic equilibrium. 
Even the most perfect \quasicrystals exhibit phasonic 
defects, and it seems very difficult to get rid of them.

Several projects have pointed up the quite important role of diffusion in 
\quasicrystals; failure to allow sufficient time for this
results in metastable states. But if experiments are performed at a 
micro- or nano-scale in order to reduce equilibration time, 
the surface energy may falsify the results. 
As most \quasicrystals are formed from the liquid state in a peritectic 
reaction (often extremely incongruent), the role of diffusion during 
single crystal preparation must be considered. As-cast states formed
under usual cooling rates and even after moderate cooling rates may be
metastable states, due to undercooling or low diffusion rates. 
Therefore, \name{L\"uck} and collaborators
have discussed the as-cast states of Al-Cu-Fe ~\cite{lueck-ascast-03}
separately from the equilibrium states~\cite{lueck-equil-03}.

\subsubsec{Probing stabilisation by non-equilibrium processing}

Regarding the low temperature stability of \quasicrystals, 
\name{Mukhopadhyay} pointed out that 
at low or even room temperature, it is extremely difficult to achieve the 
equilibrium condition in ternary and quaternary compounds, as the kinetics 
are too sluggish due to the low diffusivity: it is a frozen state.
However, we can adopt 
non-equilibrium processing techniques (such as 
electron irradiation~\cite{zhang-urban92},
ion milling, mechanical milling~\cite{mukho-mill,kupsch02,hass05},
etc.) to energise the system 
by inducing defects, thus driving it towards a stable or metastable 
configuration. 
With these techniques, we gain the help
of ballistic diffusion to overcome the problem of low thermal diffusivity. 
In fact it may
be possible to achieve the equilibrium state by suitably controlling the 
process parameters. It is therefore instructive to exhaust the possibilities 
by optimizing the experimental condition. 

Furthermore, while adopting these 
techniques there is also a possibility of obtaining a nanophase 
microstructure, in which case the overall thermal diffusivity is enhanced, 
and the diffusion distance is reduced. As a result, equilibration can be 
achieved at a faster rate even at lower or intermediate temperature range 
during the annealing/thermal treatment following after the non-equilibrium 
processing techniques.  However extreme precaution needs to be taken to avoid 
the contamination or compositional changes during these processes; otherwise 
artifacts may lead to incorrect conclusions. 

Many examples of phase transformations were found (by
e-beam irradiation/mechanical milling)
in Al-transition metal \quasicrystals  
by the groups of \name{Urban}~\cite{zhang-urban92} 
and \name{Mukhopadhyay}~\cite{mukho-mill}.
Thus it appears that these \quasicrystals are {\it not} stable at room 
temperature. Therefore, careful experiments of the kind mentioned above 
are worth pursuing 
in order to ascertain the relative stability/instability of the \quasicrystal
phase at lower temperature.  
However, what is the stable configuration in the 
(true) ground state phase is still an open question,
and needs further studies on
both the theoretical and experimental fronts.     

\subsection{Role of ab-initio calculations}
\par

\name{Widom} aserted that, in fact, 
highly accurate and reasonably efficient first-principles calculations 
are starting to give us a real possibility to know whether
there is a \quasicrystal ground state from theory 
rather than experiment.
He recommended the
`CALPHAD' method to resolve the stability of \quasicrystals.  
In brief, that means we combine experiments (e.g. on heat capacity) with
first-principles calculation (mainly at $T=0$) -- putting
together every scrap of known information -- to obtain
$G(x,T)$ at low temperatures.

The problem of \quasicrystal stabilisation is essentially a 
question of alloy phase diagram topology.  The equilibrium phase 
diagram in the $(x,T)$ composition/temperature space must be 
derivable from a free energy function $G(x,T)$.  Published phase diagrams 
drawn solely on the basis of empirical observation, without 
reference to the underlying free energy, often have
the embarrassment of details 
that are thermodynamically improbable, or even impossible~\cite{okamoto00}.
Such pitfalls are avoided through use of the CALPHAD method, 
which utilizes databases of thermodynamic functions.  This 
method suffers from a shortage of accurate data for 
\quasicrystal-forming alloys, but filling in the missing 
information can be a helpful common focus for metallurgical 
research efforts that will result in significant advances in our 
understanding.  

A newly developing approach seeks to incorporate 
data from first principles calculations~\cite{wang04}.
By a happy coincidence, first principles methods are at their most powerful 
at very low temperatures, where traditional experimental methods 
suffer from long equilibration times.  This approach can help 
refine complex alloy structures and also distinguish low 
temperature stability from metastability~\cite{mihalkovic04}.
Demanding $G(x,T)$
be consistent simultaneously with reliable experimental 
observations and with accurate calculations provides useful 
checks on both experiment and theory.  
\name{Widom} recommends that \quasicrystal researchers 
wishing to address thermodynamic stability consider CALPHAD
as an organizing principle for their efforts.
[\name{de Boissieu} has also reminded that calorimetric experiments 
are needed as one kind of input.~\cite{deboissieu-qdisc}.]

\name{Cahn} asked whether 
the melting temperature of a single compound
had ever been predicted correctly?
\name{Widom} replied that it had not, but that is not necessary. 
The proposal is not only to use first-principles calculations, but also
thermodynamic data.
The former methods are surprisingly good at low temperatures.
[The latter data are easy to obtain near melting].

\name{Cahn} asked  how the calculated total energy scales
with the order of the approximant? \name{Mihalkovi\v{c}}
replied that not enough example approximants have been computed 
to extract the scaling.

\name{Kelton} pointed out the paper of
R.~G.~Hennig {\it et al}~\cite{hennig05} which
suggested that i-TiZrNi might be an equilibrium \quasicrystal.
The best decoration model for  the \quasicrystals had
an {\it ab initio} energy lying only 3 meV/atom above the
tie-lines of competing crystalline phases.
[However, in recent {\it ab-initio} calculations of 
phase diagrams by \name{Widom} and collaborators,
which use comparatively large approximants to model
decagonal \quasicrystals,
that energy would be considered a sizeable difference.
\name{Henley}, a coauthor of Hennig and Kelton,
notes that the i-TiZrNi calculation was limited to
smallish approximants each containing only one kind of
`canonical cell' (or about 200 atoms after decoration).
That energy probably gets smaller if we take into account that 
different kinds of canonical cells are mixed, but quite possibly 
a moderate-sized approximant would be more favorable
than the \quasicrystal.]

\ack
We thank the organizing committee for arranging  this session,
and all the participants who made it lively.
We gratefully acknowledge financial support under
U.S. DOE grant DE-FG02-89ER-45405  (to C.~L.~H.) and
SNF 200020-105158 (to W.S.).

\end{document}